# Microscopic Aspects of Stretched Exponential Relaxation (SER) in Homogeneous Molecular and Network Glasses and Polymers

J. C. Phillips

Dept. of Physics and Astronomy, Rutgers University, Piscataway, N. J., 08854-8019

## Abstract

Because the theory of SER is still a work in progress, the phenomenon itself can be said to be the oldest unsolved problem in science, as it started with Kohlrausch in 1847. Many electrical and optical phenomena exhibit SER with probe relaxation $I(t) \sim \exp[-(t/\tau)^{\beta}]$, with $0 < \beta < 1$. Here $\tau$ is a material-sensitive parameter, useful for discussing chemical trends. The "shape" parameter $\beta$ is dimensionless and plays the role of a non-equilibrium scaling exponent; its value, especially in glasses, is both practically useful and theoretically significant. The mathematical complexity of SER is such that rigorous derivations of this peculiar function were not achieved until the 1970's. The focus of much of the 1970's pioneering work was spatial relaxation of electronic charge, but SER is a universal phenomenon, and today atomic and molecular relaxation of glasses and deeply supercooled liquids provide the most reliable data. As the data base grew, the need for a quantitative theory increased; this need was finally met by the diffusion-to-traps topological model, which yields a remarkably simple expression for the shape parameter β, given by d*/(d* + 2). At first sight this expression appears to be identical to d/(d + 2), where d is the actual spatial dimensionality, as originally derived. The original model, however, failed to explain much of the data base (especially polymeric relaxation, as accurately measured through concordances between multiple probes). Here the theme of earlier reviews, based on the observation that in the presence of short-range forces only d* = d = 3 is the actual spatial dimensionality, while for mixed short- and long-range forces, d* = fd = d/2, is applied to four new spectacular examples, where it turns out that SER is useful not only for purposes of quality control, but also for defining what is meant by a glass in novel contexts. The examples are three relaxation experiments that used different probes on different materials: luminescence in isoelectronic crystalline Zn(Se,Te) alloys, fibrous relaxation in orthoterphenyl (OTP) and related glasses and supercooled melts up to $1.15T_g$, and relaxation of binary chalcogen melts probed by spin-polarized neutrons (T as high as $1.5T_g$). The model also explains quantitatively the appearance of SER in a fourth "sociological" example, distributions of 600 million $20^{th}$ century natural science citations, and the remarkable appearance of the same "magic" values of β = 3/5 and 3/7 seen in glasses.

## 1. Introduction

Relaxation is a complex process, and Stretched Exponential Relaxation (SER) is exponentially complex. Exponentially complex [NPC, or non-polynomial complete, as mathematicians usually describe them] problems cannot be solved analytically or algebraically (this brilliant insight was first realized by Cantor in the 1870's, and it dominated the development of 20th century mathematics; however, these ideas are still unfamiliar to many scientists). The Scher-Lax-Phillips (diffusion-to-traps) universal minimalist model [1,2] is axiomatic, and it relies on the kind of topological and set-theoretic reasoning that has been found to provide the best approach to complex systems and large data bases. $10^{12}$

The axiomatic diffusion-to-traps model quantitatively explains "magic" stretching fractions $\beta(T_g)$ for a wide variety of relaxation experiments (nearly 50 altogether) on microscopically homogeneous electronic and molecular glasses and deeply supercooled liquids by assuming that quasi-particle excitations indexed by Breit-Wigner channels diffuse to randomly distributed traps (sinks). The basic idea behind the immobile trap model is that in microscopically homogeneous glasses nanoscale dynamical traps develop and freeze in position at $T = T_g$. Pump excitations relax by diffusion in the otherwise microscopically homogeneous medium to these fixed traps, and as time passes, their density near traps is depleted. The stretching of the exponential is then a microscopic dynamical glass memory effect. The expression $\beta = d^*/(d^* + 2)$ suggests that the low-temperature limit for $\beta$ could be 3/5 (at high T, the traps are mobile and softened, $d^* \rightarrow \infty$ and simple exponential relaxation usually occurs, $\beta = 1$), providing that the dimensionality $d^*$ of the configuration space in which the relaxation occurs is simply $d^* = d = 3$. Other values of $\beta$ imply more complex configuration spaces, which are likely to occur in glasses and deeply supercooled liquids, with their exponentially large viscosities. The model is inherently topological, because the traps cause space to be multiply connected. Surprisingly enough, there are many glasses which do exhibit $\beta(T_g) = 3/5$ (within a few %) at temperatures T near and below the glass transition temperature $T_g$. This result, in itself, is most remarkable: how often does one see such pure "magic" numbers in nature?

While the complexity of glass-forming materials and the large data bases required to explore SER limited research on this problem for many years, superfast computers with large memories have made the problems of measuring properties over wide ranges of time and temperature much more accessible. While the computer-driven transformation from scattered and unreliable data on samples of doubtful microscopic homogeneity to extremely accurate data (tested for reproducibility on multiple samples) began in the 1980's (especially with the spin-polarized data taken with the Grenoble high-flux neutron reactor), in the early 1990's this revolution occurred very rapidly. Historically it began with the first spin-polarized neutron experiments on cis-trans polybutadiene (Fig. 17 of [1]) in 1988. An example of this revolution is shown in Fig. 1, copied from Fig. 20 of [1], and presented here for readers who lack the time to study [1] properly. This example shows how dramatically the 1994-1996 diffusion-to-traps theory supported these elegant experiments, and separated them from the last of the older unreliable experiments, which had already nearly disappeared by 1993.

Although computer automated data acquisition has made relaxation experiments much easier, the results are still very sensitive to experimental design. This review discusses three very well-designed experiments (using different probes on different materials) which shed further light on the traps to which excitations diffuse, and the configuration space of dimensionality d* in which the diffusion takes place; failure of the theory to explain any one of these three experiments would threaten the soundness of the theory. Note that these three experiments used different probes on different materials, yet the 1996 axiomatic diffusion-to-traps theory not only explains their results, but reveals features of these experiments not previously identified. The discussion takes place within the general framework of glass theories that analyze how specific kinds of chemical bonding lead to efficient (maximized) space-filling without crystallization, but it also requires much detailed information on specific glassy materials. As this review builds on [1], the reader is implored at least to skim [1] (75 pages, 32 figs., 189 refs., 262 citations), in order to provide herself with an overview of this very broad and deep subject, without which the many nanoscopic comments that follow might seem to be unmotivated.

The best glass formers (slowest cooling rates without crystallization) are found in oxide network glass alloys, followed by chalcogenide alloy network glasses, but many molecular glasses are known, together with polymers and even metallic and colloidal glasses. The best glass-formers are also the ones most easily made microscopically homogeneous, an essential requirement for a microscopic theory. The characteristic feature of good glass formers is that they efficiently fill space, as reflected by constraint theory for oxide and chalcogenide network glass alloys (non-central forces) [3,4], and by free volume theories for metallic glasses (central forces) [5,6]. Constraint theory of hydrogen bonding also accurately describes the glass-forming tendencies of small molecular alcohols and saccharides [7]. The microscopic structure of good glasses often persists through the glass transition at $T = T_g$, up to a crossover temperature $T_0 \sim (T_g + T_m)/2$ or larger ($T_m$ = melting T), as reflected in the viscosity (fragility) of supercooled liquids [8]. Thus one can suppose that this structure consists of clusters percolatively connected in the glass, with the rigid percolative paths broken in the deeply supercooled liquid, and the clusters themselves decomposing only for $T > T_0$. Rigidity percolation is the characteristic feature of network glass alloys [9].

**2. Dynamics of SER**

There is a very wide range of relaxation studies, which can be implemented for a correspondingly wide range of reasons [1]. While there are some very careful relaxation studies under carefully controlled conditions, by far the most common methods are "quick and dirty", and by far the most popular of these are dielectric relaxation studies. These steady-state studies, carried out over limited frequency ranges in the presence of an applied ac electric field between plane-parallel condenser plates, can be valuable in establishing qualitative trends for τ for a range of samples or properties. As the data are taken in the frequency domain, to obtain accurate values of both τ and β, it is necessary to convert them to the time domain, and this is usually done using qualitative formulas that work reasonably well for data taken over a rather narrow frequency range (6 decades or less). (If one wants to do better, more accurate numerical programs are available [10], but these are seldom used.) In straightforward field-free temporal relaxation, studies of both polar and nonpolar molecular glasses have shown many concordances (to within 5% or better) for different materials with different probes. Unfortunately, deconvoluted values of β obtained by field-forced dielectric relaxation in the frequency domain on the same materials normally disagree typically by ~ 20% [1]. These disagreements have never been resolved; they may be caused by double layer effects [11], or by differences in relaxation in polarized compared to unpolarized states [10]; these latter differences are unavoidable, as the applied electric field must be large enough to yield a signal above thermal noise. Here we discuss temporal relaxation only, thus postponing the issue of achieving 5% agreement between theory and experiment for dielectric relaxation to another time; in fact, many recent papers on dielectric relaxation focus on approximate chemical trends only, without attempting to determine τ and β accurately [12].

In the field-free time domain matters are much simpler theoretically, as the relaxation nearly always neatly divides into two parts: an extrinsic short-time part, which reflects the history of the physical sample or the boundaries of the numerical simulation, and a longer time (asymptotic) bulk part which is fitted very accurately and economically by SER. (For example, the three-parameter SER fit is always superior to the four-parameter two-exponential fit, which is a remarkable fact in itself, first recognized by Kohlrausch in 1854. Since the fraction β is dimensionless, it appears that it must be a topological parameter dependent only on an effective dimensionality d*.) This makes it easy to separate the longer time part, and if it covers three or more decades, curve-fitting yields values of β accurate to a few %, for example, three decades in time in Fig. 4 below. (For skeptics, of whom there are still many, SER fits very accurately the neutron spin echo data on *cis-trans* polybutadiene (Fig. 17 of [1]), which span seven decades in time.) As to the question of whether or not SER is genuinely an asymptotic phenomenon, [1] cites data on a-Se based on two six-decade relaxation experiments (neutron spin echo and stress, see Table I) which span an overall time scale of $10^{12}$, with the two independent experiments concordantly giving $\beta = 0.42, 0.43$.

**3. Topological and Geometrical Models of SER**

Generally asymptotic relaxation in a glass or deeply supercooled liquid is a complex process, similar in some respects to capture of slow neutrons by nuclei. The standard Breit-Wigner method for treating the latter problem introduces separate and distinct reaction channels [13], and glassy relaxation can be similarly discussed using relaxation channels. These are supposed to represent non-vibrational degrees of freedom which can lead to relaxation. As these are not known in detail, it was long believed that SER cannot be treated quantitatively. However, in the 1970-1980's it was realized [14,15] that the central feature of SER, the dimensionless stretching fraction β, must be a topological parameter, dependent only on the effective dimensionality d* of the configuration space in which the excitation diffuses, and given explicitly by $\beta(T_g) = d^*/(d^* + 2)$. This apparently only replaces one unknown dimensionless parameter (β) with another unknown dimensionless parameter (d*). However, this is not quite correct, because it turns out that in all cases with short-range forces only, $d^* = 3$ both experimentally and in numerical simulations [1]. By the 1990's the early data base (which was littered with artifacts, based on partially crystallized and/or inhomogeneously hydrated polymers, etc.) had been superceded by many excellent asymptotic studies of carefully prepared glasses, polymers, electrolytes and supercooled liquids, as well as numerical simulations with supercomputers. When these data were collected, they exhibited simple quantitative patterns for $\beta(T_g)$, with easily justified values of d* [1,2].

At this point one can pause to take stock. Because of the exponential complexity of glasses, as reflected by superexponential increases in viscosity $\eta(T)$ as $T \rightarrow T_g$, conventional polynomial methods based on Hamiltonians, partition functions, or various random but still restricted models (including lattice percolation models) have never [1,2] reliably derived $\beta(T_g)$. Topology, however, transcends the restrictions of polynomial models and need not assume that configuration space is simply connected. Moreover, the derivations [13,14,1] of $\beta(T_g) = d^*/(d^* + 2)$ rely only on the diffusion equation and the existence of randomly distributed trapping sites (fixed in position at $T_g$) at which excitation energies are dissipated, so that the trap model (multiply connected configuration space) is truly minimal.

How successful is this minimal, yet multiply connected model? Results [1,2] from large-scale numerical simulations of simple models, experimental values based on a wide range of pumps and probes for network glasses, polymers, ionic fused salts and organic molecular glasses are listed in the executive summary given in Table 1, and it is clear by any standard that the multiply connected model is very successful indeed, in fact, nearly universal. It is amusing to note that the conclusion to [1] suggested that "in ten years the comparison between theory and experiment [may be] even closer than [now]", and that this is exactly what happened [2].

Whether or not such trapping sites could exist in glasses and deeply supercooled liquids was uncertain until the 1990's, when the successes of both simulations and a variety of experiments [1] left no doubt that the relation $\beta(T_g) = d^*/(d^* + 2)$ is often very accurate, thus fully justifying the concept of a multiply connected configuration space. Thus it is arguable that these successes by themselves provide convincing circumstantial evidence for the existence of discrete traps with abrupt boundaries, outside of which the traditional diffusion equation is valid for relaxing excitations. This conclusion is not obvious, as the glassy or deeply supercooled liquid matrix outside the rigid traps may not have the properties of a normal liquid: for example, it has been claimed that near $T_g$ the Stokes-Einstein relation between viscosity and self-diffusion coefficient fails in fragile molecular glasses such as OTP [16], as discussed in Sec. 5. In one very simple and direct case, the diffusion of Ag markers in a chalcogenide alloy glass, the traps have been identified unambiguously in realistic numerical simulations [17], thus putting to rest any doubts concerning the need for microscopic justification of the trap model.

Viscosity and self-diffusion decoupling raises questions about the validity of the universal minimalist trap model, and has led to efforts to derive SER in other ways based on plane wave scattering theory and random walks in a homogeneous continuum model [18]. Unfortunately, these efforts failed to derive the key result of the trap model, $\beta(T_g) = d^*/(d^* + 2)$. Examining their analysis, we find that the discussed liquid correlation functions are *equilibrium* functions that do not distinguish between scattering and relaxation. This distinction is important ergodically: in the trap model, excitation energies are dissipated in the rigid traps, never (in the glass) or seldom (in deeply supercooled liquids) to return to the liquid matrix (eventually they reach the sample boundaries via chains of trap states [19]). In deeply supercooled liquids the observed rapidly diffusing states are analogous to the fast ion states of solid electrolytes, which are known to follow percolative paths composed of structural units with free volumes $\mathcal{F}$ above average [20,21]. The paths are smoothly self-organized to minimize free volume differences $\delta\mathscr{T}$ between successive filamentary cells, which minimizes scattering due to density fluctuations. Large scale numerical simulations then predict ionic conductivities over 11 orders of magnitude, and show that the activation energy $E_\sigma$ for ionic conduction (analogous to a pseudogap or superconductive gap energy) scales with $\delta\mathscr{T}^{d^*/3}$ with $d^* = 1$.

It is instructive to compare the topological $\beta(T_g) = d^*/(d^* + 2)$ relation with the results of a recent simulation using an optimized hopping (not diffusive) geometrical trap model for electronic relaxation [22,23]. In the geometrical model there are four non-observable parameters, the densities of transport and trapping sites, $N_o$ and $N_T$, and the radii of these sites, a and R. The model does not give SER asymptotically at large times, but it does give curves that are accurately fitted by SER at short and intermediate times. The curves depend on two completely unknown parameters, $N_o/N_T$ and a/R, as do the fitted geometrical values of β, which is disappointing compared to the topological result $\beta(T_g) = d^*/(d^* + 2)$. Even in the simplest possible case, strongly disordered Zn(Se,Te) alloys, the meaning of $N_o/N_T$ and a/R is elusive, while d* from the diffusive diffusion-to-traps model is easily interpreted (Sec. 4).

The most subtle aspect of the diffusion-to-traps universal minimalist model is the definition of $d^* = fd$. Here d is the dimension of Cartesian scattering space, and f measures the ratio of the numbers of effective relaxation channels to their total number. In many simple cases involving short-range forces only (sphere mixtures, metallic glasses, fully cross-linked network glasses) f = 1, leading to β = 3/5. However, in more complex cases (for example, Se chains, polymers, and many electronic glasses) f = 1/2, so β = 3/7. The pervasive pattern here is that in the simultaneous presence of short- and long-range forces there are two equally weighted sets of scattering channels, only one of which is effective for relaxation. This pattern is apparent in Table I, and phenomenologically speaking it is unambiguous.

Long-range forces can arise from stress (Se and polymers) or from Coulomb forces (many electronic glasses). Recently a very elegant case (well-controlled pseudo-binary semiconductor crystals) has been studied, time-resolved luminescence from Zn(Se,Te) alloys (useful for orange light-emitting diodes), that shows the crossover from non-diffusive Debye relaxation ($\beta = 1$), first to maximally localized relaxation with $f = 1$ and $\beta = 3/5$, and then to the lower bound, $f = 1/2$ and $\beta = 3/7$ (Sec. 4). A few special cases are also known where $f = 1/3$ and $d^* = 1$; of course, these must occur because of special geometries (Sec. 5). The equal weighting of short-and long-range channels to give $\beta = 3/7$ is plausible in two respects: by analogy with the Ewald method for optimizing Coulomb lattice sums [24], and by realizing that in the nearly steady state as soon as the two channels become unbalanced, the more numerous one will lose its extra weight by relaxing faster (a kind of competitive detailed balance). Another way of seeing why the short-(long-) range forces are (in)effective for relaxation is based on the idea that long-range forces connect the diffusing particle to many traps, whereas short-range forces connect to the nearest trap. Finally, the glassy state itself can be regarded as obtained by optimized space-filling. Because space-filling itself involves a delicate balance between short-range constraints and long-range attractive forces (for instance, Van der Waals interactions), such an equalized kinetic balance between a short-range interactions (which promote relaxation) and long-range interactions (which are ineffective and do not contribute to relaxation) seems quite plausible. In this case d* = 3/2 gives $\beta_g$= 3/7, a value observed in experiments on microscopically homogenous polymers and electronic glasses. The crossover from 3/5 to 3/7 has been observed in the case of monomer (PG) to polymer (PPG) propyl glycol [2].

The question of determining d* is relatively simple so long as relaxation is dominated by quasiparticles, but glasses and deeply supercooled liquids often contain clusters, which can be regarded as part of optimized space-filling (medium-range precursors of long-range crystalline order). In Q-dependent diffraction experiments on molecular glasses the quasi-particles are identified with momentum transfers $Q = Q_1$, where the largest peak in the scattering function S(Q) occurs at $Q = Q_1$. In many molecular glasses and deeply supercooled liquids a weaker but still narrow peak (often called the first sharp diffraction peak, or FSDP, or Boson peak in Raman scattering) occurs at $Q = Q_B \sim 0.4Q_1$. The Boson peak is associated with extended topological defects (clusters) in space-filling glasses and deeply supercooled liquids, with values of $\beta(Q_B,T_g)$ qualitatively different from those of $\beta(Q_1,T_g)$. Sec. 6 extends diffusion-to-traps theory to explain SER trends and magnitudes for the Boson $Q = Q_B$ peak in deeply supercooled chalcogenide liquid alloys.

## 4. Luminescence in Isoelectronic Zn(Se,Te) Alloys

Probably the most reliable early measurements of SER in luminescence were the pump/probe photoinduced absorption studies of Cd(S,Se) nanocrystallites used in commercial optical filters [25]; similar results were obtained in fullerene $C_{60}$ films [1,2];

in both cases β(T) leveled off at low T near 0.40. Later further luminescence studies on porous Si showed β(T) again leveled off at low T near 0.40 [2]. These optical experiments on quite diverse samples strongly support the diffusion-to-traps universal minimalist model of SER [1,2] for the case of mixed short- and long-range interactions (d* = 3/2). In each case one could argue that Coulomb blockade is involved in the low-T stretching. By contrast, Si nanodots with diameters small compared to exciton radii gave d* = 3, as expected (short-range forces only) [2].

These older data are now dramatically confirmed by state-of-the-art, time-resolved studies of isoelectronic luminescence from 16 single-crystal (commercial quality) $ZnSe_{1-x}Te_x$ alloys [26]. Because many scientists are still unfamiliar with the successes achieved by SER in fitting relaxation data, an example of the 1995 data, taken from Fig. 5 of [25] is shown in Fig. 2, together with an example of the 2009 data in Fig. 3. (Also Fig. 4 of [25] shows their data fitted with two exponentials, and the superiority of the SE fit is obvious.) The latest results for the 16 single-crystal (commercial quality) $ZnSe_{1-x}Te_x$ alloys for τ(x) and β(x) are shown in Fig. 4, which has been annotated to bring out new features not discussed by [26]. Although the energy gaps at x = 0 and x = 1 are nearly equal, the alloy gap is strongly bowed downward, and the band edge shifts from 2.7 eV (x = 0) to 2.05 eV for x = 0.35. Because hole masses are much larger than electron masses, as x increases from 0, valence band edge offsets associated with Te sites localize holes on Te anion clusters, which then bind electrons, leading to bound exciton-mediated recombination. In 1993 [27] observed a continuous Localized – Extended (L-E) transition from the recombination through free and bound exciton states to the recombination of excitons localized by the compositional fluctuations of the mixed crystal in the concentration region of about x = 0.25.

[26] observed in 2008 that the narrow coherent exciton recombination band found at x = 0 with β(0) = 1 was already broadened at x = 0.005, and the exciton lifetime increased by a factor of order 100 between x = 0 and x = 0.08 (longest lifetime, most localized exciton). Meanwhile, β(x) dropped rapidly along a striking S-like curve centered on x = 0.05, to reach β(0.10) = 3/5, which is *exactly* the value expected for a fully localized (longest lifetime) state whose kinetics are dominated by short-range interactions. At x = 0.10, β(x) abruptly changes slope, decreasing more slowly to a broad, rather flat minimum with β(0.22) = 0.45, which is very close to the value 3/7 expected for a perfectly balanced mixture of short- and long-range interactions. According to [27] in 1993, x = 0.20 - 0.25 is exactly where the L-E transition takes place, effectively mixing short- and long-range forces, so again the 1996 diffusion-to-traps universal minimalist theory has *quantitatively* predicted both of the key features of this elegant 2008 experiment, as indicated in Fig. 4. Of course, at other values of x than 0.10 and 0.22, a mixture of mechanisms will determine β(x). Thus at x = 0 or 1, there is no diffusion, and there is simply exponential recombination with β = 1. Between 0 and 0.10, one has a mixture of free and bound exciton recombination, and so on. It is noteworthy that the intuitively plausible (but non-diffusive) geometrical hopping trap model, designed to discuss semiconductor impurity bands [22], is dependent on non-topological parameters, and therefore is unable to identify either of the two (not just one!) key features of Fig. 4, much less to confirm their values quantitatively. Note that these two key features involve exactly the same two magic numbers that have occurred repeatedly in Table I, which

represents the cream of the cream of all SER data taken in the last 25 years. Moreover, these two numbers are etched in granite (included in the 1996 abstract of [1]).

Looking at Fig. 4 as annotated, many scientists would readily concede that there are two special values of x that correspond to β = 3/5 and 3/7, but they would still argue that for almost all other values of x these special values are mixed with each other and with β = 1. However, it is important here to remember that $ZnSe_{1-x}Te_x$ alloys by themselves are not glasses. For most values of x the glassy behavior of luminescence in these alloys involves mixing the β = 1, 3/5 and 3/7 channels because the disorder of the alloys still takes place in a crystalline framework. As is customary in exponentially complex disordered materials, the details of the disorder are unknown. However, at the extremals of τ(x) (x ~ 0.10) and β(x) (x ~ 0.25), one can plausibly guess the key features of the disorder. For x ~ 0.10 there are small spheroidal Te-rich clusters that effectively trap holes but not electrons, causing electron-hole recombination to be extremely slow. When the clusters are large enough to localize nearly all the holes, these localized hole states will nearly fill space – in other words, near the maximum in τ(x) at x ~ 0.1, the clusters and bound holes form a conventional glassy state. With increasing Te concentration, these clusters begin to overlap, and for x ~ 0.25 the overlap is just large enough to cause a percolative Localized – Extended (L-E) transition. The percolative paths generally resemble a combination of blobs and strings, and thus mix short- and long-range interactions: these paths fill space with minimal overlap near x ~ 0.25, which defines a polymeric-like glass (with "beads"). However, actual recombination occurs only for holes in the blobs, and not on the strings. By topological symmetry, one expects the blobs and strings to have equal weight at the Localized – Extended (L-E) transition, and this gives f = 1/ 2 and d* = 3/2. Not accidentally, this gives $\beta = \beta_2 = 3/7$, just as for polymers.

The internal ordering of the alloys involves strain energies that are dependent on many bonding energies other than those involved in band-edge exciton localization and recombination. Only at the analytic *extrema* (maximum τ or minimum β) (probably dominated by band edge states) do we obtain single-channel behavior. There are two remarkable aspects of Fig. 2: (1) in spite of band edge complexities, the relaxation still follows the Kohlrausch form, and merely mixes the three β channels to generate an intermediate value of β, and (2) both f = 1, β = 3/5 and f = 1/2, β = 3/7 are found at the extrema of either τ or β. Note that such special behavior at extrema is characteristically topological in character, as it does not require analyticity, it requires only smooth connectivity. In real glasses, as shown in Table 1, the relaxation is usually dominated by a single channel with β = 3/5 or 3/7. Other values of β do occur occasionally, corresponding to mixed channels; the few known examples are discussed in [1,2]. Another such example will be discussed in Sec. 6; once the "pure" cases are understood, it becomes relatively easy to explain the mixed cases.

There is a subtle aspect of Fig. 4: the maximum in τ is narrow (f = 1), whereas the minimum in β (f = 1/2) is broad. As noted above, because space-filling itself involves a delicate balance between short-range constraints and long-range attractive forces, f = 1/2 suggests an equalized kinetic balance between short-range interactions (which promote relaxation) and long-range interactions (which are ineffective and do not contribute to relaxation). This kinetic balance should involve negative feedback, as an excess of short-range interactions will tend to relax more rapidly towards a dynamically balanced state. This stabilization mechanism is consistent with the breadth of the f = 1/2 minimum, and it suggests that departures from this minimum occur only in strongly inhomogeneous materials (such as the polymer blend examples discussed below).

## 5. **Orthoterphenyl Revisited**

Because it is commercially available in high purity, and because of its symmetrical nearly collinear three-ring planar structure, resembling anthracene, OTP is the most studied organic molecular glass former, apart from simple alcohols and sugars [7]. Already in Table 1 there are four entries for OTP, and five values for β, three close to 3/5 and two close to 3/7 (the specific choices depend on probe and (in the case of multi-dimensional NMR), pump history as well (see [1,2] for details, which are important, and which also challenge all other non-topological theories).

Optical measurements using polarized light to photoselect an orientationally anisotropic subset of probe molecules by photobleaching can also emphasize clusters. [28] studied SER in OTP in this way and found near $T_g$ that β(OTP) = 0.34 and β(anthracene in OTP) = 0.39. Values of β for five other probe molecules were also obtained, all of which satisfied $\beta \geq 0.60$. These data would appear to be widely scattered and were long considered to be mysterious, but they actually have a simple explanation in terms of the DIFFUSION-TO-TRAPS model (see Fig. 5, again annotated from [28] to bring out key features). There are simply two branches to the measured values of β, plotted in Fig. 5 as a function of the probe hydrodynamic volume normalized to the OTP host volume (upper abscissa). In OTP itself the photoselected clusters are fiber-like stacks of OTP, with fluctuations in the stacks of each molecular planar axis around the local azimuthal angle, with $d^* = 1$. Anthracene (also three collinear rings) can be fitted well enough into such a glassy stack, which is why its $\beta$ is close to that of OTP itself. The other heteromolecules (including tetracene, with an effective atomic volume intermediate between anthracene and OTP, but four rings in a square) do not fit into the fibrous stacks, and they relax independently of the OTP fibers. Starting with tetracene (four rings, $\beta = 0.60(6)$), $\beta$ increases smoothly with heteroprobe volume. This behavior is exactly what one would expect, as the volume of tetracene is close to that of OTP, while it obviously will not fit into three-ring stacks, so it should have $d^* = 3$ and $\beta = d^*/(d^* + 2) = 0.60$. As the heteroprobe volume increases, the number of degrees of freedom available for relaxation by collisions with the smaller host OTP molecules increases, and $\beta$ increases smoothly, as reported. Diffusion-to-traps theory is exact for all these OTP data; SER in OTP has recently been discussed inconclusively (without memory effects associated with dynamical heterogeneities), and it was merely suggested that $\beta$ could be 1/2 [18] (never observed!). This value of 1/2 arises when it is assumed that relaxation takes place randomly and continuously throughout the sample, rather than discretely at traps. Apparently such continuous relaxation has never been reported for the glassy state.

Of course, at the time the [28] data were reported (1995) there was no diffusion-to-traps model, and had such a model been available, it is unlikely that it would have been taken seriously. Even so, nowadays perhaps it should be, because there is now excellent evidence [29,30] for the existence of just such quasi-one-dimensional molecular fibers in OTP. First, at $T_g$ + 3K OTP exhibits anomalously enhanced (by a factor of 100) translational diffusion, as measured by surface desorption. Structurally this is most easily explained by capillary diffusion along fibrous surfaces. Note that such capillary diffusion will not invalidate the main assumption of the diffusion-to-traps model, that the bulk diffusion outside the traps is normal, because the fibers themselves are the traps.

Self-diffusion controls crystallization in OTP for most of the supercooled liquid regime, but at temperatures below $T_g$ + 10 K, the reported crystallization rate increases suddenly while the self-diffusion coefficient does not. This regime ("diffusionless crystallization", DC) has been observed in other molecular glass formers, notably ROY, currently the top system for the number (seven) of coexisting polymorphs of known structures. Some polymorphs did not show DC growth, while others did; the polymorphs showing DC growth changed growth morphologies with temperature, from faceted single crystals near the melting points, to fiber-like crystals near $T_g$. The DC mode was disrupted by the onset of the liquid's structural relaxation but could persist well above $T_g$ (up to 1.15 $T_g$) in the form of fast-growing fibers [31]. The diffusion-to-traps theory predicts that near $T_g$ $\beta$(DC mode liquids) ~ 0.35, while $\beta$(not DC mode liquids) ~ 0.6. This is a strong prediction for the seven coexisting polymorphs of the ROY system, and if confirmed, it would establish a more detailed topological criterion for DC growth than is provided by current free volume models [30].

The reader will have noticed that (including the values in Table I) we have seven different values for $\beta$(OTP), including the two branches of Fig. 5. These occur as 3/5 (four times), 3/7 (twice) and 1/3 (once). It is impossible to explain these multiple values (much less their narrowly clustering around the same "magic" fractions that occur for many other materials!) by using any geometrical model based only on isotropic free volume concepts and spatial heterogeneity. In particular, the attempt to extend the Gibbs equilibrium nucleation model (a spherical model, independent of dimensionality) to discuss SER by using modified Gaussian fluctuations [31], while qualitatively interesting, is fundamentally unsound and in fact misses the essential topologically asymptotic nature of SER in much the same way as the spherical geometrical model [23]. [31] is inadequate in many different ways, but the notion that the glass transition is somehow or other governed by "random" (Gaussian) statistics is perhaps its most unsatisfactory feature.

Another very interesting example of SER is provided by NMR spin-lattice relaxation and stimulated echoes in ice II, which behaves quite differently from ice I [32]. (This in itself is surprising, as ice I is hcp, and ice II is ccp, an apparently small difference. However, note that the phase diagram of ice closely resembles that of silica, an archetypical glass-former. Specifically ice I is analogous to tridymite, and ice II to cristobalite.) The SER of defects in ice I showed large $\beta \sim 0.8$-$1.0$, while that of defects in the high-pressure phase ice II have $\beta = 0.6$, a difference the authors found puzzling. By adding these data to Fig. 5 of this paper, one is led to a plausible model. The defects in as-grown ice I are large-scale relative to a single molecule (volumes about three times the host volume), whereas those in high-pressure ice have collapsed to nearly the same volume, the collapsed "point" volumes presumably caused by high pressure internal stresses. Note that this application uses OTP (a large, planar organic polycylic molecule) to establish a length scale for small, tetrahedral $H_2O$. Thus it appears that SER can be used to establish a topological length scale for glassy defects that is independent of host morphology (this is a question much-discussed in model simulations [33]); in itself, this is remarkable, as topology is generally supposed to describe only scale-independent effects, but here it succeeds in handling length scales, presumably because of the compact nature of glassy configurations.

## 6. Rigid Cluster Relaxation in Supercooled Chalcogenide Alloy Liquids

Given the configurational complexity of glasses and supercooled liquids, one might well have supposed *a priori* that $\beta_g$ would have no microscopic significance, and would vary inexplicably from one material to another, and from one probe to the next, even for the same material. Surprisingly enough, this view is still widely held [34], in spite of the numerous successes of the universal minimalist diffusion-to-traps model, both systematic [1] and predictive [2]. Those successes, however, are not readily achieved for data from poorly prepared (microscopically inhomogeneous) samples (for example, partially crystallized or hydrated polymers, where relaxation can be dominated not by intrinsic diffusion, but by nucleation kinetics).

Diffraction is a technique which at first seems to raise special problems, as $\beta = \beta(q)$, and one can expect that $\beta(q) \to 0$ as $q \to 0$ (low probe energy), while $\beta(q) \to 1$ for $q \gg Q_1$. Spin-polarized neutron scattering experiments on molecular glasses (OTP and other examples listed in Table I), as well as several numerical simulations, all agree that $\beta(Q_1)$ [the largest peak in the scattering function $S(Q)$ occurs at $Q = Q_1$] is in excellent agreement with non-diffraction probes [specific heat, ultrasonics, light scattering], so that the choice $Q = Q_1$ is remarkably successful in converting the scattering data to quasi-particle kinetics. So far this concordance of the diffraction value of $\beta_g(Q_1)$ with $\beta_g$ in other temporal relaxation experiments is merely a plausible experimental fact, so here a possible path towards justifying it is suggested. The path involves constraint theory, which identifies intact constraints starting from $r = 0$ in radial distribution functions in network glasses [3, 4] and continues until enough constraints have been accumulated to exhaust the available degrees of freedom. In Q space this same procedure starts at large Q and continues towards smaller Q. Even in complex molecular glasses, where the appropriate steric hindrance constraints are not known, this process will obviously stop at or near $Q = Q_1$, so that in glasses $\beta(Q_1)$ should be in excellent agreement with $\beta_g$ obtained from non-diffraction probes [specific heat, ultrasonics, light scattering].

It is clear that a different meaning must be attached to relaxation in supercooled liquids of excitations associated with the Boson (or first sharp diffraction) peak at $Q = Q_B \sim 0.4Q_1$ ; if Kohlrausch relaxation was the oldest (150 years) unsolved problem in science, then explaining $\beta(Q_B)$ must be the most difficult problem for theory (or at least condensed matter theory). Here we discuss the first data on $\beta(Q_B,T)$ obtained by scattering spin-polarized neutrons for six binary chalcogen melts [35]. The relaxation here refers to some kind of cluster, and one might well suppose that here (at least) diffusion-to-traps theory will finally fail. The present theory does not succeed in predicting the data on $\beta(Q_B,T)$ to a few % accuracy (as it did for $\beta_g(Q_1)$), but it does explain systematic chemical trends. Moreover, these unique experiments test the significance of the effectiveness factor f in d* = fd in a wholly unexpected way. Historically such unique experiments can provide strong support for heuristic or axiomatic model theories.

Boson peaks, either in diffraction, Raman scattering, or vibrational spectra (either theoretical or from neutron scattering) are the characteristic signature of extended glass clusters [36]. In molecular metallic, or colloidal glasses (such as orthoterphenyl, OTP) boson peaks are associated with cages around vacancies or miscoordinated (relative to the crystal) sites [37,38], while the boson peak in metallic glasses is modeled elastically in terms of local structural shear rearrangements [39]. These clusters are precursors of crystallization, and their structure involves weak interactions compared to the stronger interactions discussed in constraint theory (see above), which determine the composition dependence of the glass-forming tendency.

Chalcogenide network glass alloys have traditionally provided the best examples of First Sharp Diffraction Peaks and Boson peaks [36,40]. There the FSDP greatly increases in strength from Se chains to predominantly tetrahedral $GeSe_2$, where it corresponds to the Ge-Ge interplanar spacing in a locally layered structure. Detailed studies of the composition dependence of $Q_B$ and the amplitude of the FSDP show that it is weak in Se and strong in the layered compound compositions $As_2Se_3$ and $(Si,Ge)Se_2$. At intermediate cation concentrations there is an abrupt crossover from Se inter-chain dominated spacings to interplanar spacings for cation concentrations near 0.1, where the height of the FSDP also shows an abrupt change in slope [41].

Chalcogenide alloy glassy networks exhibit a Boson-like peak even in their vibrational densities of states measured by neutron scattering, called "floppy modes" [42,43]. With increased cation cross-linking the network stiffens, and in mean-field theory (constraint theory) this gives a stiffness transition [3,43]. Well-homogenized glasses self-organize and form percolative networks near the stiffness transition, giving rise to an exponentially complex intermediate phase (IP) [44]. Internal network stress is reduced by a large factor in the IP, as measured by Brillouin scattering for both chalocogenide alloys [45] and sodium silicate alloys [46]. Most of the rigid modes have condensed to form percolative paths [47,48], but some may have condensed to form small clusters which contribute to the FSDP.

The structural effects of internal network stress observable by diffraction are small, but they were evident in the reduced width of the FSDP in the average over 4 closely spaced compositions in the $Ge_xSe_{1-x}$ IP window $0.20 \leq x \leq 0.25$, compared to 14 closely spaced compositions outside the window (Fig. 9 of [49], reproduced and annotated here as Fig. 6). Moreover, as pointed out by Prof. G. Lucovsky (private communication), the PDF width shows a sharp second minimum near x = 0.30, where Raman spectra (1977) had show that a stress-relieving ethane-like structure $(Se_{1/2})_3$-Ge-Ge-$(Se_{1/2})_3$ is present [50,51]. There is an interesting technical point here. To compare Extended XrAy Fine Structure width data, which contain subcomponents of the Pair Distribution Function data, with the PDF width data for the FSDP, the authors of [49] combined the former assuming that there were no Ge-Ge contacts in the network. As can be seen from Fig. 6, this had the effect of erasing the sharp minimum at x = 0.30. Thus even had the Raman data not been known, in principle careful study of the differences between the PDF and combined EXAFS data would have suggested the onset of the ethane-like structure $(Se_{1/2})_3$-Ge-Ge-$(Se_{1/2})_3$ near x = 0.30. Moreover, since the effect shows up much more strongly in the FSDP than in Raman scattering, one can conclude that the ethane-like structure dominates the FSDP cluster for x above x = 0.30.

While the interpretation of their data given by the authors of [49] was overly conservative, it is true that diffraction is not the only tool for studying the FSDP clusters. With spin-polarized neutron scattering one can study the relaxation of clusters near Q = $Q_B$ in chalcogenide (Se,As, Ge) alloy melts [35]. The data here are complex and involve high temperatures (the lowest temperatures studied at different compositions range from 1.2 $T_g$ to 1.6 $T_g$), where the clusters are much weaker than near $T_g$. However, the data contain some interesting structural features and one spectacular trend.

First it is helpful to picture the effects of adding cross-linking As or Ge to Se chains. The difference between As and Ge cross-linking is simply that As will be asymmetrically bonded (twice to one chain, once to the other), whereas the Ge bonding will be symmetrical (twice to each chain). Thus the Ge cross-linking clusters preserve the set-theoretic equivalence of the Se chains, and the activation energies $E_a(\langle r \rangle)$, where $\langle r \rangle$ is the average coordination number in the alloys, are a nearly linear function of $\langle r \rangle$ from $\langle r \rangle = 2.0$ (Se) up to $\langle r \rangle = 2.4$ in $GeSe_4$, but the more curvilinear line drawn through the As alloys does not pass through Se (see Fig. 7). (The same feature is evident in the temperature coefficient $C_\beta$ defined below.) This means that the relaxation of cross-linking clusters is more complex in the As alloys than the Ge alloys, which will reappear in chemical trends in $\beta(Q_B)$.

The most striking feature of the data of [35] is their most remarkable discovery that $\beta(Q_B)$ = exp($-C_B(x)/T$), where x diffusion-to-traps relation d* = fd, we expect the effective fraction f of relaxation channels to be small for large clusters (most degrees of freedom will not be effective in relaxation of large clusters which have only a few pinning points, which again helps to explain why the chemical trends in $\beta(Q_B)$ are more complex for the asymmetrically pinned As than the symmetrically pinned Ge alloys). With $f_B << 1$, $\beta(Q_B) = f/(2/d + f)$ is nearly proportional to f. One would expect the effective fraction f of relaxation channels to be thermally activated, which is exactly what [35] discovered. This is a totally unexpected and so quite satisfying way of confirming the physical meaning of f, and actually proving that f is an effective fraction of relaxation channels.

The chemical trends in $C_B(x)$ are informative. Because Se chains in the melt are long [1], $C_B(x)$ should be smallest in Se, and increase with $<r>$, as observed. In fact, the reported values of $\beta_B$(Se) are close to 0.43 near $T_g$, which is nearly the same value as $\beta(Q_1)$ in Se (Table I). In other words, for unalloyed Se clustering effects are small, and the relaxation of the long chains is nearly independent of Q. However, as soon as cross-linking begins ($Ge_{08}Se_{92}$ in inset of Fig. 4 of [35]), $\beta(Q_B)$ and f drop sharply (by about 1/3). Now one no longer has merely competition between short- and long-range forces (as in Se), but direct competition between effective and ineffective relaxation modes of clusters.

At these large values of $T/T_g$ the separation between $Q_1$ and $Q_B$ relaxation channels is not sharply defined in β(Q), but it is still perceptible where the clusters are best defined and f is smallest ($AsSe_3$ in Fig. 3 of [35] near Q = $1.8A^{-1}$). In their unpublished data (private communication) for $GeSe_4$ near $1.55A^{-1}$ there is a break in slope suggestive of incipient phase separation at 720K (~ 1.5 $T_g$ ). For the same value of <r> f is smaller for the asymmetrically pinned As than for the symmetrically pinned Ge alloys, consistent with the activation energy differences discussed above.

**7. Special Cases: Proteins and One-Dimensional Geometries**

This paper has by choice studied unambiguous examples of SER. At the opposite limit nonexponential relaxation of proteins in water and other solvents is complicated by competing relaxation and memory effects of solvent and solute. Nevertheless, recent studies of dielectric relaxation of myoglobin (Mb) in water and glycerol over wide temperature ranges showed consistent patterns when compared to Mossbauer and neutron scattering data [52]. These patterns imply that internal Mb relaxation is enslaved to "$\beta_h$" relaxation of its hydration shell. One component of charge transfer relaxation after photodissociation of MbCO was identified (their Fig. 6C) and assigned a relaxation fraction (which they denoted by *p*) of 0.65. The diffusion-to-traps model suggests that this component is associated with short-range interactions, as 0.65 is close to 3/5 (see Fig. 3 here for molecules in OTP). The data also exhibit a long-time tail, which according to the patterns surveyed here and previously [1,2], probably corresponds to a smaller relaxation fraction near 0.4 associated with long-range Mb interactions. SER is a characteristic signature of glasses, and its definitive appearance in a protein strengthens the analogies between proteins and glasses. Note also that a continuum "downhill" model of protein folding does not yield SER, but one involving discrete barriers (which will trap the protein diffusion-to-traps model of SER is an important adjunct to protein transition state analysis [54].

The special value $\beta = 1/3$ ($d^* = 1$) for spin glasses [1] is now well-supported by at least two experiments [55]. Another way to produce relaxation in a $d^* = d = 1$ (literally one-dimensional) geometry is through relaxation of transient drain currents of polymer channel transistors, whose decay is well fitted by a stretched exponential function with $\beta = 0.28\text{-}0.29$ [56]. (Had the channel been ideally linear, $\beta$ would have been 0.33, but with rough spots ($d^* = 0$) the measured value is reduced. Note that the reduction of this value from 0.33 can be used as a very simple and accurate way to measure channel quality.)

The size-limiting mechanism ([2] and Sec. 4 here) that explains why luminescence in small nanodots decays more rapidly (β = 3/5) than in nanodots with dimensions large compared to exciton radii (β = 3/7) has also been identified as affecting dopant stability [57]. Note that this size mechanism is intrinsic and does not require defects or dangling bonds on the nanodot edge. Also self-bleaching in photodarkened Ge-As-S amorphous thin films exhibits SER with β = 3/7 (not surprising, since photodarkening certainly involves competition between long wave length darkening and molecular scale rebonding) [58], so that the topological list in Table I is by no means exhaustive. As previously noted [1,2], the dominant mechanism for polaron relaxation in large quantum dots involves long-range forces through longitudinal optic phonons, as confirmed by recent quantum-dot intersublevel transitions in the terahertz range [59].

**8. Rigor of Models**

In [1,2] and here it has been assumed that [14,15], which treats SER as asymptotically exact, is correct, but this conclusion is not universally accepted: more recently, numerical simulations using hopping (not diffusion) [22] have led to the conclusion that SER occurs only over short and medium times, and that the key relation $\beta = d/(d + 2)$ is not valid. The methods used by [14,15] are analytical, and derive from techniques introduced by many distinguished theorists (starting with I. M. Lifshitz in 1964), while [22]'s numerical simulations appear to be unambiguous, so there is a real puzzle here, which [1,2] and here have chosen to resolve in favor of Lifshitz *et al*. by appeal to experiment.

If we look closely at [15] and [22], we immediately recognize a crucial difference: [15] treats the carrier motion as continuously diffusive in d dimensions, whereas [22] uses a model involving hopping between randomly distributed discrete points. It appears that the discreteness of the hopping (d = 0) model radically alters the nature of the relaxation, moving it from the asymptotic long-time to the near-time regime. One could argue that the hopping model [22] is suitable for the luminescence experiments discussed in Sec. 4, but then the extremal values of 3/5 and 3/7, identified there using [15] and $\beta = d^*/(d^* + 2)$, would remain unexplained. This suggests that in the extremal cases a continuum approximation is valid, but in fact the time intervals are essentially the same for all compositions studied. The conclusion appears to be that in all cases the continuum (with traps!) model is correct, and that SER is a long-time property, as has been assumed by almost all workers for many years [1].

## 9. Separating the Wheat from the Chaff (aka Quality Control)

Prior to the appearance of [1], an uncritical survey of much older (mostly pre-digitalized) $\beta_g$ data appeared [60]. At that time many believed that the ability of stretched exponentials to give better fits to data was merely a peculiar and inexplicable accident, but as we have already seen (Figs. 2 and 3), as sample homogeneity and data improved, so did the accuracy of SE three parameter fits! There were ~ 70 $\beta_g$ values in [60], containing heterogeneous polymer blends (for instance, resins) and many other poorly characterized, microscopically inhomogeneous materials [61,62], and including many values obtained by dielectric relaxation (20% scatter from temporal values measured by many other better understood methods), a contact method involving crude transformation from narrow frequency ranges to the time domain. Not surprisingly, these archaic (almost prehistoric) data (including some from the 1950's and 1960's) are widely dispersed, and this is often cited as a reason for doubting that it is possible to construct a meaningful theory of SER. In fact the 70 $\beta_g$ values span a range from 0.2 to 1.0, which is far different from the strongly concentrated and concordant modern values shown here in Table I.

The risks inherent to collecting data (especially polymer data) uncritically have been brought out by recent studies of SER in bipolymer blends [59]. These were interpreted in terms of a microscopically inhomogeneous model with local compositions defined over varying spatial scales. The relaxation spectra could not be fitted with a single SER function. Complex statistical analysis disentangled the relaxation and revealed bifurcated $\tau(Q)$ and $\beta(Q)$ patterns, implying complex cross-relaxation between the regions with differing local compositions. These experiments dramatically emphasize the importance of maintaining microscopic homogeneity in relaxation studies. Nor is that all: molecular dynamics simulations show that a given monomer in a miscible polymer blend experiences broad distributions of both connectivity driven self-concentrations and thermodynamically controlled intermolecular concentration fluctuations. The distribution of self-concentrations is particularly important in the dilute limit, where intermolecular concentration fluctuations should be absent. These conclusions rationalize recent literature results that report the apparent self-concentration determined in the dilute limit surprisingly depended on the blend partner [61]. These polymer results might even discourage one from thinking that progress could be made in understanding impurity relaxation in a molecular glass, but the results obtained here (Fig. 5) for OTP show that molecular glasses are much simpler than polymers.

More generally, even in microscopically inhomogeneous materials, SER can be explained by stress relaxation [63], assuming the absence of internal cross-relaxation between regions of differing local compositions. However, microscopic homogeneity and the extremal properties associated with maximized space filling are essential to the clustering of $\beta_d$ analyzed by the diffusion-to-traps model.

A detailed discussion of sample quality for each of the older values may be unnecessary, as even the modern examples (Fig. 1 - Fig. 3) show improvements in accuracy by factors of order 10-100, but how do we know that even the modern values of $\beta_g$ can be trusted? Suppose that the older $\beta_g$ values were uniformly distributed, then 0.05 wide boxes centered on 3/5 and 3/7 would contain about 5 older values each. If the 50 modern values were equally uniformly distributed, they would have about 3 modern values in each box, with a standard deviation of order 1.7. However, Table I contains 17 entries in the 3/5 box (eight sd's above random), and 15 entries in the 3/7 box (seven sd's above random). Another way of looking at these statistics is to ask how many modern values of $\beta_g$ should be found outside the three 0.05 wide $\beta_g$ boxes, assuming that $\beta_g$ is uniformly distributed between 0.33 and 1.0, given that 50 lie within the three 0.05 wide $\beta_g$ boxes? The answer is 200, yet the recent papers [18,22] that assume nothing is known about this distribution were not able to cite even one such example. There is even more to this story, as the modern data not only crowd into three "magic" boxes, but there are also internal concordances in each box separately with respect to materials and pump/probes that are fully consistent with the axiomatic diffusion-to-traps theory. These concordances would bring an additional statistical factor of order $3^{50}$ into the probability against the theory being correct by accident; collecting all these factors, the chance that the modern values are so concordant by accident are about 1 in $10^{50}$.

At this point it should be clear that indeed, there has been an enormous change in standards of materials preparation in the last 20 years, amounting to a revolution that actually surpasses the one in computer science. Indeed, although data on poorly prepared samples are still quite common, and probably always will be (see also Fig. 1), nowadays studies of such samples usually provide only estimates of $\tau$, and seldom venture to report inadequately supported values of $\beta_g$; nearly all the ones that have been reported lie in the diffusion-to-traps "magic" boxes. However, could a subjective element still account for the appearance of "magic" values of 3/5 and 3/7 for $\beta_g$? The answer to this question has recently appeared in a way that no one expected.

**10. Topology and the Web of Twentieth Century Science**

Scientific communication is an essential part of modern science: whereas Archimedes worked alone, Newton (1676) acknowledged that "If I have seen a little further, it is by standing on the shoulders of Giants." How is scientific communication reflected in the patterns of citations in scientific papers? How have these patterns changed in the 20th century, as both means of communication and individual transportation changed rapidly, compared to the earlier post-Newton 18th and 19th centuries?

Scale-free network statistics, with power-law exponents, dominate many data bases, including the World Wide Web, the Internet, metabolic networks and organic chemistry [64]. Early informetric studies of citation patterns of "only a few" million citations identified such power law patterns, which were interpreted in terms of a cumulative advantage model (analogous to properties of equilibrium materials near a critical point; in everyday terms, the rich get richer) or preferentially wired networks. However, more recent studies have shown a mixed distribution, with power laws prevailing at larger numbers of citations and stretched exponentials giving better fits for smaller numbers [65]. A recent very large scale study of 20th century citations [66] found that stretched exponentials give better fits to citation chains for both low and intermediate citation levels $n < n_1$, where $n_1 \sim 40$ earlier in the century, and $n_1 \sim 200$ later. That region accounts for 95% of the citations, and it, not the scale-free high end, probably objectively represents the essential features of the citation chain. The recent very large scale study of 20th century citations included 25 million papers and 600 million citations [66].

The startling result found in the informetrics study [66] is shown in Fig. 8. Not only do the intermediate distributions exhibit SER, but they also bifurcate in 1960, with stretching exponents β that match (within a few %) the $\beta_g = (\beta_{g11}, \beta_{g22}) = (3/5,3/7)$ values predicted in 1996 in the abstract of [1]: 3/5 holds < 1960, while 3/7 holds > 1960! To be more precise, the fitted values are $\beta_c = (\beta_{c11}, \beta_{c22}) = (0.57,0.47)$.

Of course, scientific culture is far from microscopic homogeneity, with significant differences between North and South America, Europe, Africa and Asia, and this could account for the reduction in the splitting of the informetric citation values relative to physical values found in glasses. It is very interesting that if we regard $\beta_c$ and $\beta_g$ as the diagonal elements of $\beta_{ij}$ matrices (i,j =1,2) in citation and glass spaces respectively, then $\beta_{c11}+\beta_{c22}=\beta_{g11}+\beta_{g22}$, in other words, the traces of the $\beta_c$ and $\beta_g$ matrices are equal to within 1%.

The space-filling character of glasses is both general and difficult to formulate explicitly, except in the case of network glass alloys, where constraint theory has been very successful, especially in describing the glass transition [9,42,44]. However, a direct connection between constraint theory (which involves short-range forces only) and glass matrix relaxation space is not apparent. Moreover, the predicted results for $\beta_{g11}$ and $\beta_{g22}$ are so general that constraint theory does not appear to be the only way to connect them to $\beta_{c11}$ and $\beta_{c22}$.

The citation process can be compared to luminescence of electron-hole pairs, with electrons representing citing papers and holes representing cited papers. In general, radiative electron-hole recombination is inefficient: most pairs do not recombine radiatively, and similarly most papers are not cited in other papers. As we saw in Sec. 4, there are two stable cases of SER, the first arising from a maximum in $\tau$, corresponding to $\beta_{g11}$ = 3/5 (short-range forces only), and the second corresponding to $\beta_{g22}$ = 3/7 (mixed short-and long-range forces). Thus it is not surprising to find a parallel matrix bifurcation in $\beta_{c11}$and $\beta_{c22}$.

Continuing the matrix analogy, we see that to explain the reduction in the splitting of the informetric citation values relative to physical values found in glasses, we must assume that while the glass matrix can be taken as diagonal, the citation matrix is anti-Hermitian, with identical imaginary off-diagonal values $\beta_{c12} = \beta_{c21}$ = ai, with a << 1. Anti-Hermitian matrices arise in the context of a new and very efficient way to calculate two-electron excited states of many-electron atoms and molecules, without doing configuration interaction, by using an anti-Hermitian contracted Schrodinger equation (ACSE) [66,67]. In view of the well-known parallels between temperature-dependent dynamics for the partition function Z ($dZ/d\gamma = -HZ$, with $\gamma = 1/kT$) and the Schrodinger equation for the time-dependent wave function $\psi$ ($d\psi/dt = iH\psi$), this suggests a way of understanding both citation distributions and glassy relaxation. Given the indubitable concordances between the citation and glassy matrices, such parallels are attractive.

Within the ACSE one calculates two-body charge density matrices that can be said to be the classical analogue of citer-cited pair interactions. The ACSE is an optimized reduction of the enormously complex problem of quantum configuration interaction. In principle the variationally optimized two-body charge density matrices are related to the three- and four- body charge density matrices, which in turn are related to still higher order charge density matrices. Moreover, because interference effects are not included, these classical correlation functions in general do not correspond to actual wave function solutions of the Schrodinger equation itself. Nevertheless, it has been found in many examples that restricting the higher order corrections to the leading anti-Hermitian third-order term yields excellent results, at least an order of magnitude better than obtained from the full (Hermitian + anti-Hermitian) third-order term [64, 65]. The anti-Hermitian term satisfies what is known as the Brillouin condition, which means that it does not couple coherently (is out of phase) to errors in the optimized two-body charge density matrices. Classically speaking, the anti-Hermitian interaction corresponds to damping of the pair interaction by a third particle; alternatively, since the original two-body charge density matrices are already optimized, the anti-Hermitian third-order term is most effective, because it is out of phase from the Hermitian two-body term. Apparently such damping does not occur in glasses, probably because of the nearly optimized space-filling condition. It does occur in citation space (albeit only weakly); it admixes a small fraction of additional long-range interactions into the "pure" short-range case, and vice-versa. It may be that this damping represents a loss of information between the details of the short-range interactions, and the generality of the long-range ones.

What factor in scientific culture could correspond to optimized space filling by glasses, and what corresponds to short- and long-range forces? The physical distinction between short- and long-range forces may have a social analogy. Scientific collaboration occurs on multiple levels, from details to general ideas and motivation. Direct collaboration requires frequent contacts of individuals, and appears quite analogous to short-range interactions (nearest and sometimes second nearest neighbors, but seldom third neighbors) in glasses. The nature of this glassy environment depends on its size. In particular, the apparent dimension of the environment is likely to be d = 3 (latitude, longitude, and time displacements of the citations since article publication). Since there is a strong tendency to cite papers that one knows best, and these are often the papers of one's colleagues, and possibly their closest colleagues, the citation number for a given paper will be limited largely to this circle. This corresponds to $\beta_1 = 3/5 \sim 0.57$, which gives the best fit to low and intermediate citation levels prior to 1960.

Because the detailed aspects of a research paper are best known only to this small space-time circle, only the more general aspects will give rise to a larger number of citations. [Aside: in all likelihood, these 600 million citations probably form the best possible data base for studying the rational behavior of such small space-time circles, which are called "affinity groups" by sociologists. Of course, it remains to be seen whether or not all affinity groups are rational.] After 1960, scientific conferences became popular and grew steadily in size, for a wide variety of practical reasons – an upsurge in funding for science after Sputnik (supporting all expenses paid international travel for scientists), rapid moderation of commercial jet fares (the first successful commercial jet, the Boeing 707, was delivered in waves, with the first two delivery waves peaking in 1960 and 1967), etc. At scientific conferences one encounters a wider range of ideas in a much less detailed format, which corresponds to long-range interactions, and appears to give a broader distribution of citations. The wider range of ideas can survive longer, the paper can continue to accumulate citations, and because $f \sim ½$ the diffusive interactions affect only details, while leaving the larger ideas intact and still capable of accumulating citations. Here it is quite striking that the balance between short- and long-range effects on relaxation in glasses should be so closely echoed (similar values of f!) in the competition between narrower and wider influences on citation patterns.

The hardest question has been left for last. Few scientists will be surprised to learn that scientific communications have a diffusive character, or that the "trappy" events that terminate a paper's citations are randomly distributed. While the history of scientific discoveries is often written so as to make them appear inevitable, those closest to the discoveries have often remarked on the surprisingly large part played by accident, a famous case being the discovery of penicillin in bread mold.

The most difficult question then is, why does the distribution of scientific citations follow the predictions of SER ***in glasses*** so well, and what does this tell us about science and scientific research? Quite naturally [1,2] emphasized the universality of SER in microscopically homogenous glasses, including polymers. However, current research is showing that polymer blends [61,62] often exhibit multiple SER (not single SER, as shown in Fig.1 here for citations, and in [4] for many microscopically homogeneous glasses, including polymers). To explain such successful predictions there must be internal factors in science that give scientific research a glassy character, and specifically correspond to microscopic homogeneity. A possible candidate for glassy character is the search for novelty, which expands to fill the accessible space densely, just as a glass, subject to stronger direct forces, is compacted by weaker residual forces. A possible candidate for microscopic homogeneity is the insistence that scientific theories be validated by experiments – this is what ultimately limits branching balkanizations such as Lysenkoism.

This leaves one last question. Granted that careful scientific research leads to SER, why does it lead either to $\beta_1$ or to $\beta_2$? The answer to this question may be contained in Fig. 2. There we notice that although there is a wide range of values of $\beta$, the two common values are metastable extrema, either in $\tau$ or in $\beta$. The characteristic feature of extrema is their stability, which in turn favors consensus. While scientists seek originality, they also attempt to place their new results in the context of established results, which is most readily accomplished with papers that have stabilized their informational contexts. Thus (of course only in retrospect!) the appearance of SER in citation patterns, and even its rather striking change in character around 1960, appears obvious.

At the risk of making an obvious yet controversial point, it is worth noting that, although power-law scaling with accurate exponents is observed very widely [64], at present there exist very few examples where those exponents can be explained or calculated, and they are only rarely predicted. In fact, the only exception known to this author is the evolutionary exponents found in the solvent-accessible surface areas of protein amino acids [68]. Thus the predictive successes of the trap model [4-6] in explaining SE distributions in glasses and 20th century citation patterns acquire a special significance.

**11. Anomalous Diffusion**

The conventional diffusion equation in normal liquids is based on short-range **r** forces and is written

$$\partial\rho/\partial t = D\nabla^2_{\mathbf{r}}\,\rho \qquad (1)$$

In glasses with short-range forces the diffusion to traps model leads (15) to $\beta = d/(d+2)$, which is in excellent agreement with earlier results (1-4) and those discussed here. How should (1) be generalized to describe the case of mixed short **r**- and long-range **R**-forces? The physical content of (1) is that the spatial fluctuations on an **r** scale drive the temporal diffusion. In the mixed **r R** case, the spatial fluctuations on an **r** scale can drive spatial fluctuations on an **R** scale, which in turn can drive temporal diffusion. This picture is the dynamical relaxation aspect of the "multiple length scale" approach to inhomogeneities (68), and it leads to the anomalous diffusion equation

$$\partial\rho/\partial t = D'\nabla^2_{\mathbf{R}}\nabla^2_{\mathbf{r}}\,\rho \tag{2}$$

At this point one can use dimensional analysis on both (1) and (2) to argue that if (1) gives $\beta = d/(d+2)$, then (2) gives $\beta = d/(d+4)$, and with $d = 3$, $\beta = 3/7$. One can also reach this conclusion in a more elaborate way by assuming that $\rho(\mathbf{r},\mathbf{R}) = \rho_l(\mathbf{R})\rho_s(\mathbf{r})$, with $\rho_l(\mathbf{R})$ scaling as $L^{-fd}$, and with $\rho_s(\mathbf{R})$ scaling as $L^{-(1-f)d}$. Then (2) can be solved applying the methods of (15) twice to give

$$I(t) \sim \exp[-(t/\tau_1)^{\beta l}]\,\exp[-(t/\tau_2)^{\beta s}] \tag{3}$$

with $\beta l = fd/(fd + 2)$ and $\beta s = (1-f)d/((1-f)d + 2)$. The fastest relaxation (maximum entropy production [(69)]) at long times will occur percolatively along paths for which $f = 1/2$ and $\beta l = \beta s = d/(d+4)$. The extra "push" given to short-range fluctuations by long-range forces explains the anomalously enhanced (by a factor of 100!) surface diffusion reported for fragile molecular glass formers like OTP (29,30). Anomalous surface diffusion is routinely observed for polymers [70]. Even macromolecules diffusing through small pores exhibit [71] enhanced diffusion, with $\beta \sim 0.8$. Looking at Fig. 5, we see that the enhanced $\beta$ reflects a typical dimension of the flexible macromolecules of order twice the pore size. Almost all observations of anomalous diffusion are qualitative, whereas the result $\beta = 3/7$ is both universal and quantitative.

## 12. Conclusions

We have seen that the diffusion-to-traps model is capable of drawing extremely detailed conclusions about the SER shape parameter $\beta_d$ for a wide variety of microscopically homogeneous glasses; indeed, each of the three physical examples discussed here is representative of the most modern state of the art for its own class, and each could well stake a claim to being "best ever' in terms of providing rich insights into the complex relaxation processes first defined more than a century ago by Arrhenius.

Perhaps most satisfying is the progress that modern experiments on high-quality samples have made possible in identifying microscopic configuration coordinates in the exponentially complex context of glasses and deeply supercooled liquids. Far from being a limitation [22] of the diffusion-to-traps theory, microscopic configuration coordinates are the substantive basis for connecting the abstract mathematical (asymptotic) aspects of SER to real experiments, as has been shown here for three elegant physical examples and one totally unexpected sociology-science one. To the author's knowledge, there is no other example in nature of such "magic" numbers being predicted so reliably and reproducibly, in such a wide range of high quality materials. All the results discussed here and below are new, and represent new insights into the nature of pump/probe relaxation in a wide variety of materials. It is striking that no other theory (such as [18, 22]), has been able to predict such results for even one specific material, much less 50 separate and distinct experiments. When one adds to these "merely 50" examples the cultural patterns of 600 million 20$^{th}$ century citations, the universality of SER becomes one of the most impressive discoveries of theory in recent decades.

The comparison between universal glass relaxation and citation patterns is instructive for both subjects. We noted that while 95% of citations follow SE patterns, the top 5% follow power-law distributions. Recently neutron scattering has shown that in metallic glasses the packing of atomic clusters is self-similar [70]. The medium-range order has the characteristics of a fractal network with a dimension of 2.3, and is described by a power-law correlation function over the medium-range length scale. The break-down of chemical packing models beyond a length scale of a few clusters appears to be universal for metallic glasses, and can be compared to the length scales of molecular glasses shown in Fig. 5 here. One can argue that this power-law distribution reflects the fact that more rapid quenching is required to prevent crystallization in metallic glass alloys, compared to network glasses.

The exponent 2.3 seen for metallic glasses is similar to the average of two (in- and out-) connectivity exponents (2.1 and 2.7) found for organic chemicals and the World-Wide Web networks [71]. There are many exponents $\gamma$ for different chemical distributions [72], with range $1.5 \leq \gamma \leq 3$. This range is topologically suggestive of fd, with d = 3 and $1/2 \leq f \leq 1$; the short- and long-range interaction model may also be applicable to power-law scaling. Most self-similar network-growing power-law models predict $\gamma \geq 3$, which is unsatisfactory. However a recent model with local and global connectivity gives $2 \leq \gamma \leq 3$, which is a substantial improvement over local connectivity models only [73]; although only a toy model, this model is analogous to the physical diffusion-to-traps SER glass model.

I am grateful to M. D. Ediger, W. C. Chou and D. L. Price for correspondence, and to J. R. Chelikowsky for drawing my attention to [57].

| Method | Material | β(exp) | β(theory) | d* |
|---|---|---|---|---|
| Num. Simula. | Spin glass | 0.35 | 0.33 | 1 |
| Num. Simula. | Binary soft sphere | 0.62 | 0.60 | 3 |
| Num. Simula. | Coord. Alloy | 0.59 | 0.60 | 3 |
| Num. Simula. | Axial quasiX | 0.47 | 0.473 | 9/5 |
| Num. Simula. | Polymer | 0.59 | 0.60 | 3 |
| Stress Relax. | Se | 0.43 | 0.43 | 3/2 |
| Spin-Pol. Neutron | Se | 0.42 | 0.42 | 3/2 |
| Method | Material | β(exp) | β(theory) | d* |
| Stress Relax. | Se-As-Ge | 0.61 | 0.60 | 3 |
| Stress Relax. | $B_2O_3$ | 0.60 | 0.60 | 3 |
| Stress Relax. | $Na_2O{\cdot}4SiO_2$ | 0.63 | 0.60 | 3 |
| Stress Relax. | PVAC | 0.43 | 0.43 | 3/2 |
| Stress Relax. | PMA | 0.41 | 0.43 | 3/2 |
| Spin-Pol. Neutron | PB | 0.43 | 0.43 | 3/2 |
| Spin-Pol. Neutron | PVME | 0.44 | 0.43 | 3/2 |
| Spin-Pol. Neutron | PH | 0.44 | 0.43 | 3/2 |
| Spin-Pol. Neutron | KCN | 0.58 | 0.60 | 3 |
| Ultrasonic | KCN | 0.40 | 0.43 | 3/2 |
| Brillouin | KCN | 0.47 | 0.43 | 3/2 |
| Specific Heat | PG | 0.61 | 0.60 | 3 |
| Specific Heat | Glycerol | 0.65 | 0.60 | 3 |
| Specific Heat | OTP | 0.60 | 0.60 | 3 |
| Brillouin | OTP | 0.43 | 0.43 | 3/2 |
| Multidim NMR | OTP | (0.59,0.42) | (0.60,0.43) | (3,3/2) |

| Method | Material | β(exp) | β(theory) | d* |
|---|---|---|---|---|
| Specific Heat | Salol | 0.60 | 0.60 | 3 |
| Ultrasonic | Glycerol | 0.60 | 0.60 | 3 |
| Brillouin | BD | 0.58 | 0.60 | 3 |
| Brillouin | HT | 0.60 | 0.60 | 3 |
| Brillouin | Salol | 0.60 | 0.60 | 3 |
| Spin-Pol. Neutron | OTP | 0.62 | 0.60 | 3 |
| Photo-Ind Absorp | $C_{60}$ | 0.40 | 0.43 | 3/2 |
| Photo-Ind Absorp | Cd(S,Se) | 0.40 | 0.43 | 3/2 |
| Carrier Relaxa. | a-Si:H | 0.44 | 0.43 | 3/2 |
| Carrier Relaxa. | porous Si | 0.4 | 0.43 | 3/2 |
| Photon Correl. | PG | 0.61 | 0.60 | 3 |
| Photon Correl. | PPG | 0.43 | 0.43 | 3/2 |
| Stress Relax. | PPG | 0.42 | 0.43 | 3/2 |
| Luminescence | Si nanodots | 0.57 | 0.60 | 3 |

Table I. Executive summary of values of stretching fraction $\beta_g$ and topological dimensionality d* below $T_g$ previously discussed at length in [1,2]. There are 38 examples here. The present paper adds four more examples for Zn(Se,Te) alloys and OTP, as well as a discussion of the FSDP clusters in supercooled chalcogenide alloy liquids, and several more examples in the postscript, bringing the total close to fifty. Although many $\beta_g$ values for polymer blends are quoted in older collections [59], none of these were included in [1,2] and here, as it is well known that simple SER does not occur in polymer blends, where neither nor $\tau$ nor $\beta_g$ are well defined [50,51]. More details, including grouping by materials and pump/probe, can be found in [1,2].

## Figure Captions

Fig. 1. Dependence of $\beta$ on vinyl fraction in polybutadiene. The older 1991 data are compared with 1996 spin-polarized neutron data (references are given in [1]). Note that in 1994 the diffusion-to-traps theory contradicted the poor 1991 data, and correctly predicted the 1996 data.

Fig. 2. Representative relaxation data at T = 20K from two-color pump-probe luminescence in semiconductor nanocrystallite doped glasses (Fig. 5 of [25]). The scatter in the β(T) values [25] suggests uncertainties of order 10-20% in β.

Fig. 3. The SE fit to the luminescence data ([26] and private communication) is accurate to about 0.2% for β (parameter b).

Fig. 4. Annotated and redrawn data [26] on luminescence in isoelectronic Zn(Se,Te) alloys. The peak in relaxation time τ occurs near x = 0.10, where there is a break in dβ/dx. These are the maximally localized states, which behave as quasi-particles subject to short-range forces only, with β =3/5. The localized-extended transition occurs at x = 0.22, where the long-and short-range forces are equally weighted, and β = 3/7. The β fitting errors are about ten times smaller than the size of the data points.

Fig. 5. Relaxation in OTP [28] exhibits two branches of β, as discussed in the text, corresponding to d* = 1 and d* = 3+. The original figure [28] has been annotated and the defect points for Ice I and II added [32]. Of course, for the molecular points the volume (upper abscissa) is normalized to that of OTP, while for the defect Ice points it is normalized to the volume of $H_2O$. The figure shows how β in OTP can be used to establish a length scale for defects in Ice, starting from the data of [28] for β of heteromolecules in OTP, which is a remarkable demonstration of the power of SER analysis to transcend materials.

Fig. 6. The measured widths of the FSDP are represented in this annotated figure from [41] by solid (black) circles; the other symbols refer to EXAFS data, which have been deconvoluted neglecting Ge-Ge contacts. Note that the measured widths of the FSDP exhibit structural features discussed in the text that are absent from the EXAFS data processed by neglecting Ge-Ge contacts.

Fig. 7. Relaxation activation energies $E_a$ of the FSDP in As- and Ge-chalcogenide alloys [35]. The smooth lines show that the Ge alloy values extrapolate smoothly to pure Se, but the As alloy values do not extrapolate smoothly to pure Se (see text).

Fig. 8. Evolution of the number of papers n with a number of citations $\tau$ (or T) as a function of the decade in which the paper was published. The distributions for the decades up to 1960 are well fitted by an SER with $\beta = 0.57$, while the decades after 1960 are fitted with $\beta = 0.47$. The crossover at 1960 is unambiguous [63].

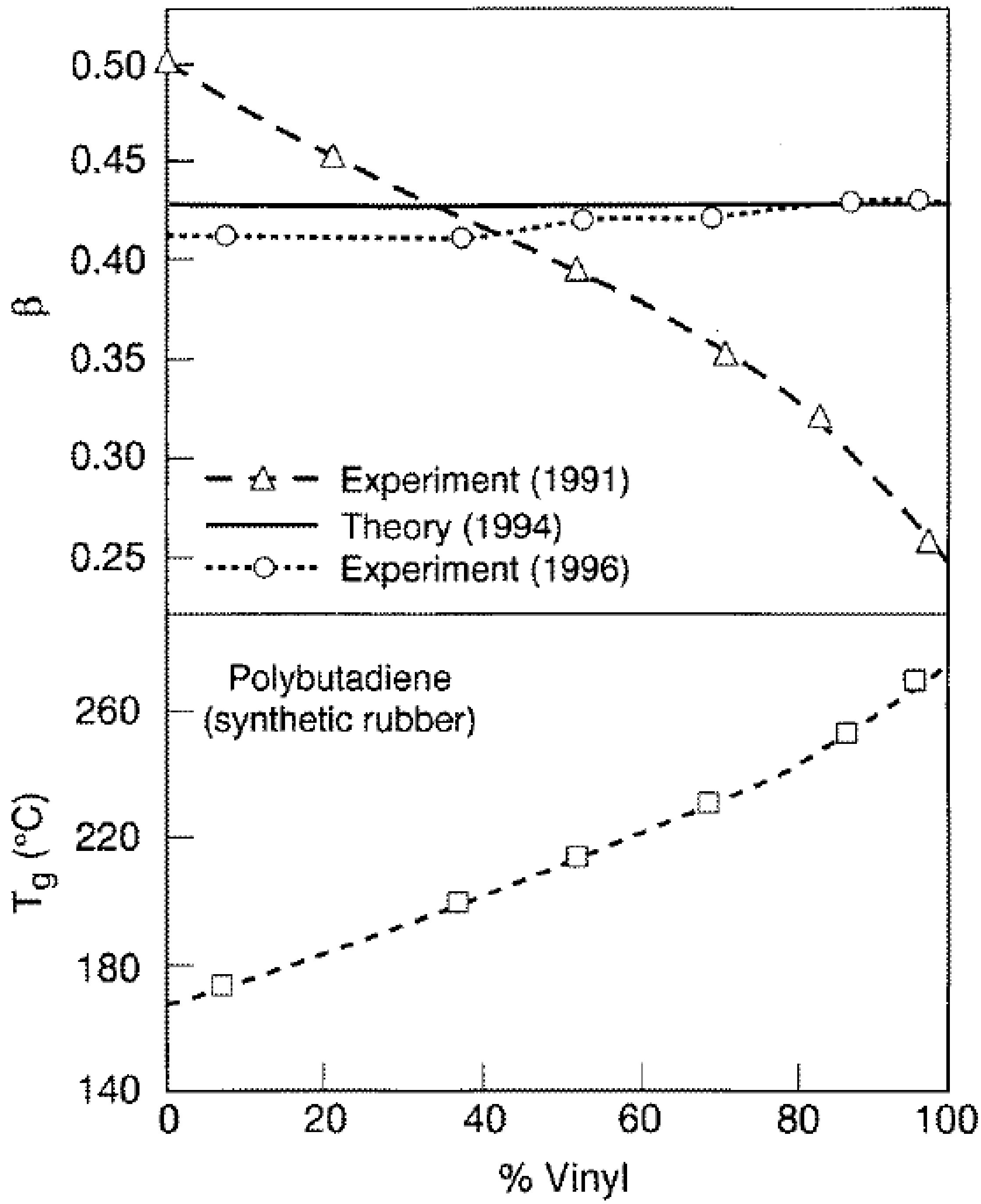

0.50
0.45
0.40
0.35
0.30
0.25
β
Experiment (1991)
Theory (1994)
Experiment (1996)
Polybutadiene
(synthetic rubber)
260
220
180
140
$T_g$ (°C)
0
20
40
60
80
100
% Vinyl

Fig.1.

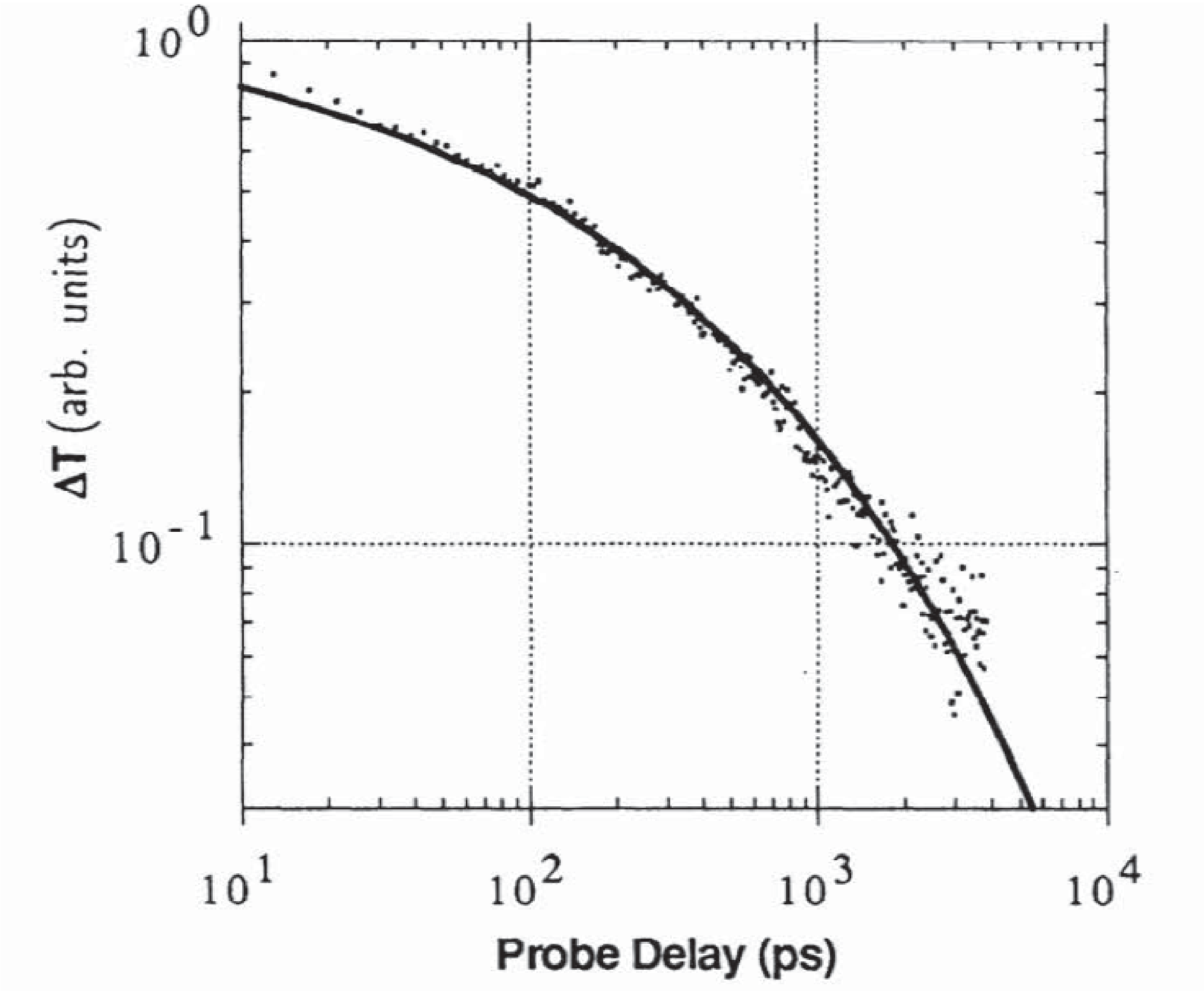

Fig. 2.

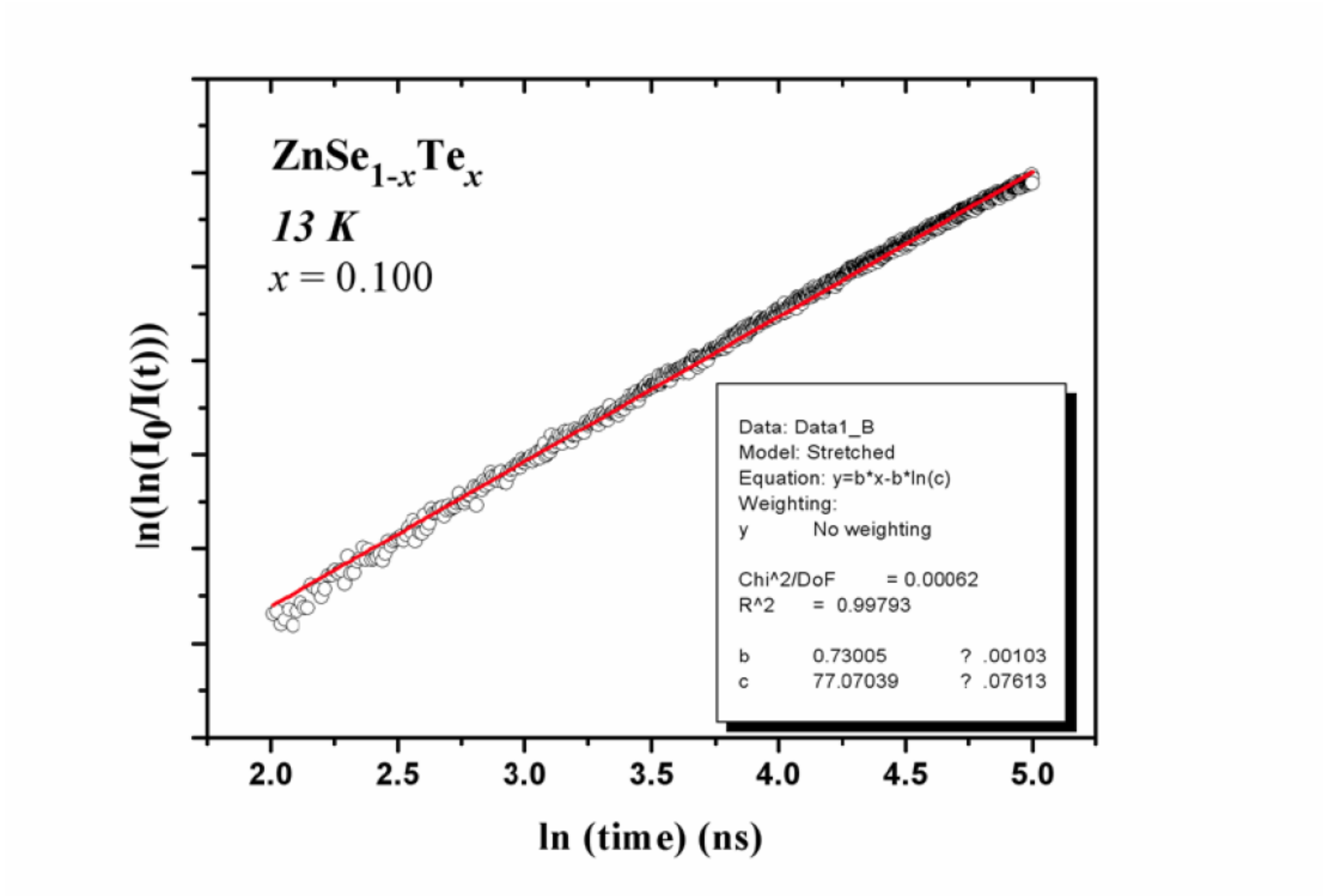
ZnSe1-xTex
13 K
x = 0.100
ln(ln(I0/I(t)))
ln (time) (ns)
2.0
2.5
3.0
3.5
4.0
4.5
5.0
Data: Data1_B
Model: Stretched
Equation: y=b*x-b*ln(c)
Weighting:
y No weighting
Chi^2/DoF = 0.00062
R^2 = 0.99793
b 0.73005 ? .00103
c 77.07039 ? .07613

Fig. 3.

.

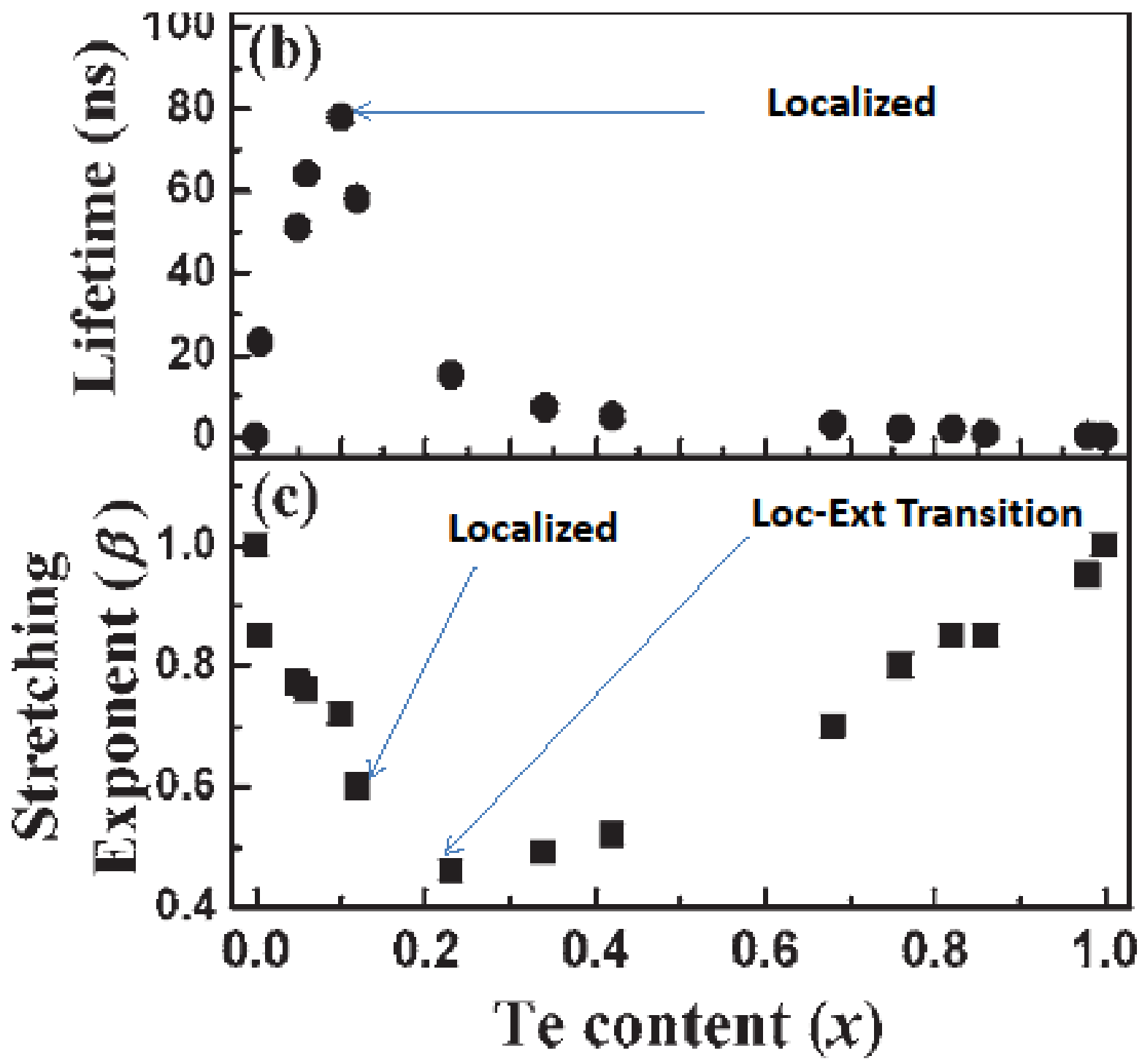
(b)
Localized
Lifetime (ns)
100
80
60
40
20
0
(c)
Localized
Loc-Ext Transition
Stretching Exponent (β)
1.0
0.8
0.6
0.4
0.0
0.2
0.4
0.6
0.8
1.0
Te content (x)

Fig. 4.

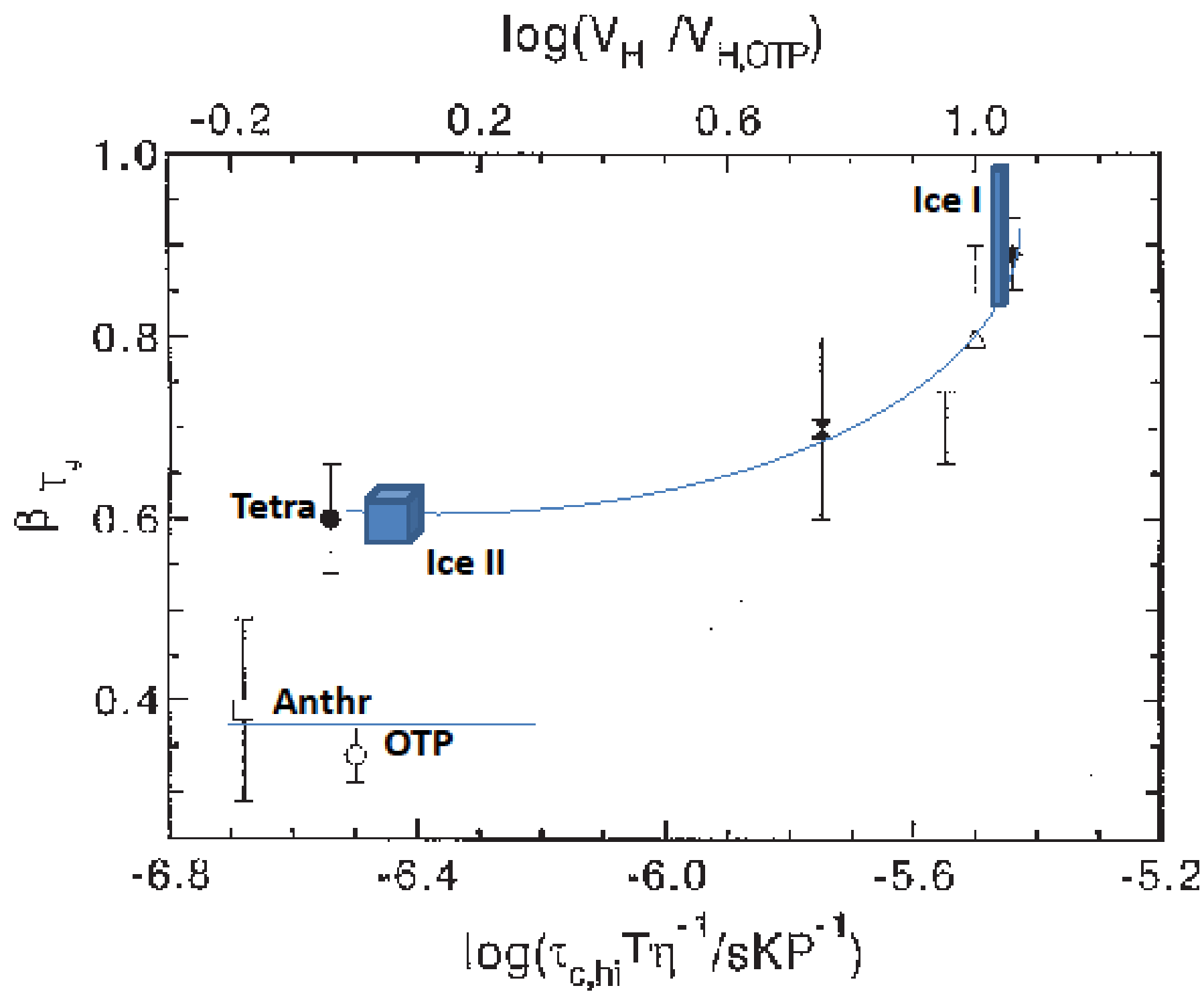

log(V_H /V_H,OTP)
-0.2
0.2
0.6
1.0
1.0
0.8
0.6
0.4
β τ
Ice I
Tetra
Ice II
Anthr
OTP
-6.8
-6.4
-6.0
-5.6
-5.2
log(τ_c,hi Tη^-1/sKP^-1)

Fig. 5.

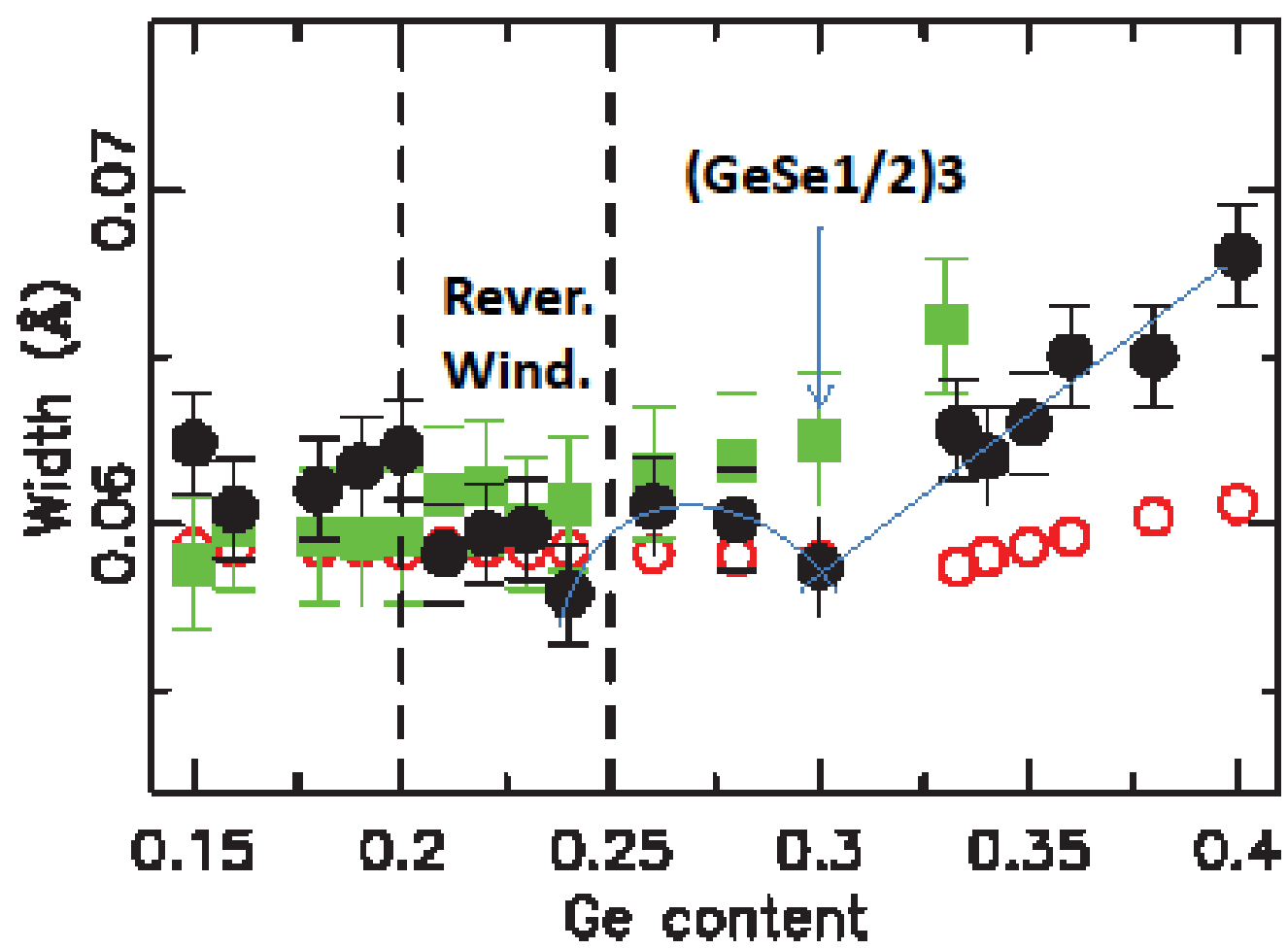
Width (Å)
0.06
0.07
Rever.
Wind.
(GeSe1/2)3
0.15
0.2
0.25
0.3
0.35
0.4
Ge content

Fig. 6.

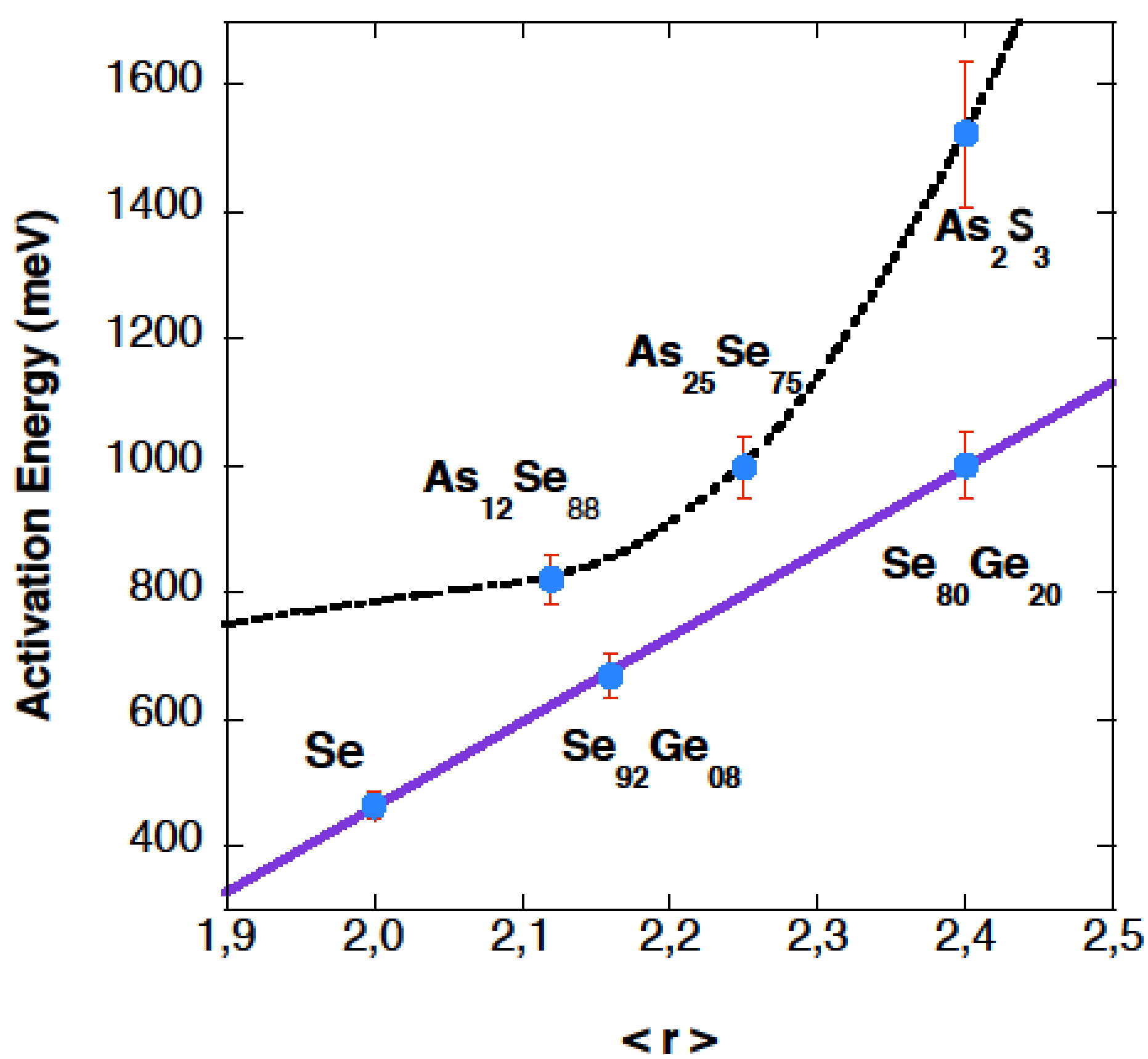
1600
1400
1200
1000
800
600
400
Activation Energy (meV)
1,9
2,0
2,1
2,2
2,3
2,4
2,5
< r >
As2S3
As25Se75
As12Se88
Se80Ge20
Se92Ge08
Se

Fig. 7.

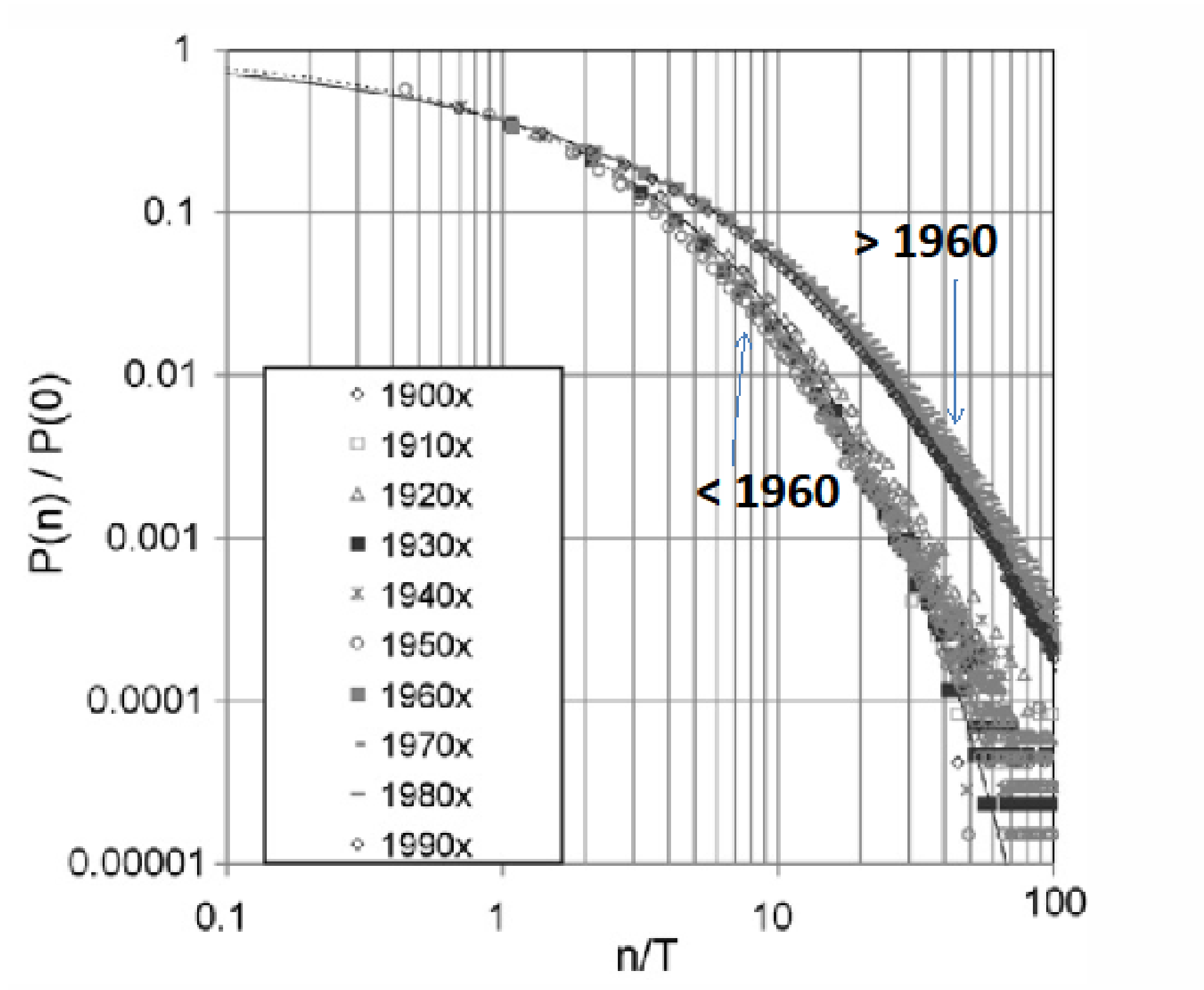
P(n) / P(0)
n/T
1
0.1
0.01
0.001
0.0001
0.00001
0.1
1
10
100
> 1960
< 1960
1900x
1910x
1920x
1930x
1940x
1950x
1960x
1970x
1980x
1990x

Fig. 8.